	\newif\ifpdf
	\ifx\pdfoutput\undefined
	\pdffalse 
	\else
	\pdfoutput=1 
	\pdftrue
	\fi
\documentclass{appolb}
\usepackage{epsfig}
\def\TT{$t\bar{t}$~}
\def\MET{$\not$E$_T$}
\def\ET{E$_T$}
\def\Journal#1#2#3#4{{#1} {\bf #2}, #3 (#4)}

\def\NPB{{\em Nucl. Phys.} B}
\def\PLB{{\em Phys. Lett.}  B}
\def\PRL{\em Phys. Rev. Lett.}
\def\PRD{{\em Phys. Rev.} D}

\ifpdf
\DeclareGraphicsExtensions{.pdf, .jpg, .tif}
\else
\DeclareGraphicsExtensions{.eps, .jpg}
\fi
\begin{document}
\pagestyle{plain}

\newcount\eLiNe\eLiNe=\inputlineno\advance\eLiNe by -1
\title{TOP QUARK PHYSICS AT THE TEVATRON\\
RESULTS AND PROSPECTS%
\thanks{Presented at Cracow Epiphany Conference on Heavy Flavours, 
January 3-6, 2003, Krakow, Poland.}%
}
\author{Krzysztof Sliwa\\
\address{Tufts University, Department of Physics and Astronomy\\
Medford, Massachusetts 02155,USA\\e-mail: 
krzysztof.sliwa@tufts.edu,krzysztof.sliwa@cern.ch}}
\maketitle
\vspace{-0.5cm}
\begin{abstract}
\noindent
Measurements of the top quark mass and the \TT and single top production cross sections, obtained by CDF and D0 Collaboration at the Tevatron, are presented.
The methodology of CDF and D0 top quark analyses and their underlying assumptions are
summarized.
The CDF and D0 top mass averages, based on about 100 pb$^{-1}$ of data from collisions of $\rm p\bar{\rm p}$ at $\sqrt{\rm s}$ = 1.8 TeV
collected by each experiment in Run-I, and obtained from measurements in several channels,
are M$_t=176.1\pm 4.0 (stat) \pm 5.1 (syst)~ GeV/c^2$ 
and M$_t=172.1\pm 5.2 (stat) \pm 4.9 (syst)~GeV/c^2$, respectively.
The combined Tevatron top quark mass is M$_t=174.3 \pm 3.2 (stat) \pm 4.0 
(syst)~GeV/c^2$. The CDF measurement of the \TT cross section (assuming 
M$_t=175~GeV/c^2$) is $\sigma_{tt}=6.5 \pm ^{1.7}_{1.4}$ pb, and  the D0 value
(assuming M$_t=172.1~GeV/c^2$) is $\sigma_{tt}=5.9\pm 1.7$ pb. In 
anticipation of the increased amount of data in Run-II, prospects are presented. The fact that top quark analyses
 are among the best windows to new physics beyond the Standard Model
is emphasized.

\end{abstract}

\section{Introduction}
\smallskip
The top quark was 
expected in the Standard Model (SM) of electroweak interactions
as a partner of the b-quark in a SU(2) doublet of the weak isospin, in the third
family of quarks. Searching for the top quark was the primary 
physics objective in Run-I.
The first published evidence appeared in a CDF~\cite{CDF94} 
paper in 1994, and its observation (discovery) was reported by CDF~\cite{CDF95}
and D0~\cite{D095} in the same issue of PRL in 1995. Both experiments have
finished their analyses of Run-I data for some time now, and only a few new 
results on top quark are
presented in this paper. A summary of the top quark's mass and the 
\TT and single top production cross section
measurements is presented. A closer look at the analysis techniques used and a perspective view on top quark physics after its first 7 years (or so) is the subject of
this paper.

In anticipation of Run-II, in which the number of reconstructed \TT events is 
expected to be at least 20x larger than in Run-I, the question of whether all 
available results are consistent with the simplest hypothesis, adopted by both CDF and D0 experiments in 
Run-I,  that data contains just the \TT events and the Standard Model backgrounds is re-visited.

\section{Signatures of \TT Pair Production and single top production}
\smallskip
At Fermilab Tevatron energy, $\sqrt s$ = 1.8 TeV, the dominant production
mechanism for top quarks is $t {\bar t}$ pair production by a
quark-antiquark or gluon-gluon initial state via the strong interaction; for top quark masses above
M$_t \approx 120~GeV/c^2$ the $q \bar{q}$ fusion dominates.
Assuming the Standard Model decays, there are three classes of final states, 
all with two b-quarks jets:
i) {\it {di-leptons}}, when both W decays are leptonic, with 2 jets and missing
transverse energy (\MET), $BF \approx 4/81$ for $e,\mu$ final states;
ii) {\it {lepton+jets}}, when one W decays leptonically and the other into 
quarks, with 4 jets and \MET, $BF \approx 24/81$ for $e,\mu$;
iii) {\it {all-hadronic}}, when both W's decay into
quarks, with 6 jets and no \MET, $BF \approx 36/81$.
The backgrounds to the possible $t\bar{t}$ signal coming from the
$W$+jets
process of QCD have been feared to be much larger than the top 
signal and must be addressed. The QCD multijet production background is even more important for all-hadronic final states, in which both W bosons decay into quarks.

\smallskip
The two dominant electroweak processes leading to a single top quark production are: a) s-channel W* production and its subsequent decay into t and b quark/antiquarks, leading to a final state with a W and two b-quark jets; b) t-channel W-gluon fusion process, leading to a final state with a W and two jets but only one of them being due to the b-quark.

\section{Top Mass and Cross Section Measurements: Methodology}
\subsection{Measurement of Cross Section}\label{subsec:cross}
\smallskip
The techniques used in CDF and D0 are variations of simple event counting.
Both experiments follow identical steps: i) identify events with the 
expected top 
signature; ii) calculate the expected SM backgrounds; iii) count events
above the expected backgrounds; iv) apply corrections for the acceptance, 
reconstruction inefficiencies and other biases. This paper reports on
measurements of the \TT
pair-production cross section and the single top
production cross section.
\smallskip
One should remember two facts: i) it is {\it assumed} that the selected sample
of events contains just the \TT events and the SM background; this is the
simplest and the most natural hypothesis since the top quark is expected in 
the SM;
ii) some of the acceptance corrections are strongly varying functions of the
top quark mass, M$_t$, and, consequently, the value of the measured cross 
section depends on
the value of M$_t$, which has to be determined independently.

\subsection{Direct Measurement of Top Mass}\label{subsec:massdirect}
\smallskip
All mass measurement techniques used by CDF and D0 assume that each event in the
selected sample contains a pair of massive objects of the same mass (\TT quarks)
which subsequently decay as predicted in the SM. 
Information about the kinematics
of the event is used in a variety of fitting techiniques. A one-to-one mapping
between the observed leptons and jets and the fitted partons is assumed.
\smallskip
Again, there are two things to remember: i) it is {\it assumed} that the selected sample
of events contains just the \TT events and the SM background; 
ii) the combinatorics of the jets-lepton(s) combinations (only one of many
possible combinations is correct) adds to the complexity of the problem.

\subsection{Indirect Measurement of Top Mass}\label{subsec:massindirect}
\smallskip
Precision measurements of various electro-weak parameters, whose values
depend on M$_t$ indirectly (via radiative corrections), are compared with the corresponding values
predicted by the theoretical calculations in the consistency checks of the SM. 
Data from
LEP-I, LEP-II, SLD, CDF, D0, $\nu$-scattering results and other precision experiments,
including or excluding the direct measurements of the 
top quark mass, can be used
to yield the most likely top quark mass, consistent with the predicted values
of the measured electroweak observables. Results are model dependent, as one
has to assume a particular theory (e.g. SM or Minimal Supersymmetric Standard Model) to make such comparisons
possible. An additional uncertainty comes from the unknown Higgs boson mass, 
which also enters into calculations of the radiative corrections. 
The results of such global
fits are summarized  in section 8, where the question of overall
consistency of all electroweak measurements is examined within the framework of the Standard Model.

\section{Top Mass and Cross section measurements.}

\subsection{Direct Searches}
\smallskip
All CDF and D0 searches impose stringent identification, selection and 
transverse energy (\ET)  cuts on leptons and jets to minimize the SM and 
misidentification backgrounds. Except for di-lepton samples, in which
backgrounds are expected to be small, various techniques of tagging b-quarks
are employed to reduce backgrounds. ``Soft-lepton" tagging is used by both CDF and D0, and the 
secondary vertex tagging, which uses a silicon vertex detector (SVX), by CDF.
D0, not equipped with a SVX, makes much greater use of various kinematic
variables to reduce backgrounds. The largest SM background is the QCD W+jets
production. Both CDF and D0 use VECBOS~\cite{Giele} 
calculations to estimate shapes of the
background distributions due to this process. Presently available samples of
the top event candidates are small, and the top cross section and mass 
measurements
are still dominated by the statistical errors. This will no longer be true in 
Run-II.

\begin{table}[h]
\caption{\footnotesize Results of D0~\cite{D0sear} and CDF~\cite{CDFsear} direct top searches.}
\begin{center}
\footnotesize
\vspace{-0.2cm}
\begin{tabular}{|l|c|c|c|c|}
\hline
channel		& D0 sample & D0 background & CDF sample & CDF background \\
\hline
di-lepton	& 5	& 1.4$\pm$0.4	& 9	& 2.4$\pm$0.5 \\
\hline
\begin{minipage}{1.0in}
\begin{flushleft}
lepton+jets \\ SVX tagged
\end{flushleft}\end{minipage} 
			&&		& 34	& 9.2$\pm$1.5 \\
\hline
\begin{minipage}{0.8in}
\begin{flushleft}
lepton+jets \\soft-lepton tagged
\end{flushleft}
\end{minipage} 
		& 11	& 2.4$\pm$0.5	& 40	& 22.6$\pm$2.8 \\
\hline
\begin{minipage}{0.8in}
\begin{flushleft}
lepton+jets \\topological cuts
\end{flushleft}
\end{minipage} 
		& 19	& 8.7$\pm$1.7	&	&  \\
\hline
all-jets	& 41	& 24.8$\pm$2.4	& 187	& 142$\pm$12 \\
\hline
$e \nu$		& 4	& 1.2$\pm$0.4	&	&  \\
\hline
$e \tau, \mu \tau$	&& 		& 4	& $\approx$ 2 \\
\hline
\end{tabular}
\end{center}
\end{table}

\subsection{Mass Measurement in lepton+jets channel}

\smallskip
In the lepton+jets and all-jets final states there is
sufficient number of kinematical constraints to perform a genuine fit.
The measured lepton and jets' four-momenta are treated as the corresponding 
input lepton and quarks' four-momenta in the kinematical fits. Leptons are 
measured best, jets not as well (in Run-I better in D0 than in CDF), while the \MET~has
the largest error.
In the lepton+jets channel one may, or may not, use \MET~as a starting point for
the transverse energy of the missing neutrino. In their published analyses
both CDF and D0 use \MET. 
However, one should remember that even in a genuine \TT event only one out of  a large number of possible lepton-jet combinations is the correct one. Quite often the incorrect combination may 
yield a false solution with a comparable or even a better $\chi^2$ than the correct combination. It is very important how the likelihood is defined; it may include just the conservation of energy and momentum, or some dynamical factors reflecting the expected production and decay characteristics of  \TT events.
CDF defines four independent samples of lepton+jets events, and
measures the top quark mass in each of them. The results are summarized in
Table~\ref{tab:cdfljet}, and presented in Figure 1.

\begin{figure}[h]
\begin{center}
\ifpdf
\includegraphics*[width=5in, height=4.5in, trim=0 0 -10 0]{cdf_mass_fits_final}
\else
\psfig{figure=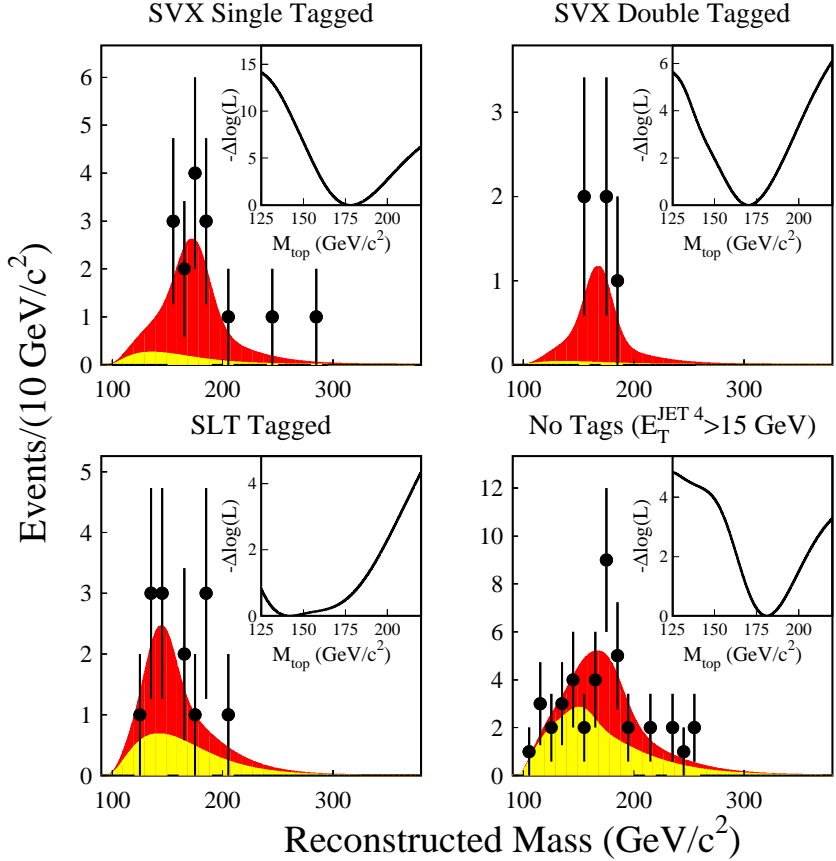,height=4.5in}
\fi
\caption{{\footnotesize CDF measurements of the top quark mass in lepton+jets
samples.}}
\end{center}
\end{figure}

\noindent
The dominant systematic uncertainties (in GeV/c$^2$) are: jet energy 
measurement (4.4); final state radiation (2.2); initial state radiation (1.8);
shape of background spectrum (1.3); b-tag biases (0.4); parton distribution
function (0.3), yielding the total systematic error of 5.3 GeV/c$^2$.

\noindent
The combined CDF result from the lepton+jets channel is:
\medskip
\begin{center}
	M$_t$=175.9$\pm$4.8(stat)$\pm$5.3(syst) GeV/c$^2$.
\end{center}

\begin{table}[h]
\caption{\footnotesize CDF top mass measurements in lepton+jets samples.\label{tab:cdfljet}}
\begin{center}
\footnotesize
\begin{tabular}{|l|c|c|c|}
\hline
subsample	& N & expected background fraction & M$_t$ (GeV/c$^2$)\\
\hline
\begin{minipage}{1.0in}
\begin{flushleft}
SVX double \\ tagged
\end{flushleft}
\end{minipage} 
		& 5 & 5$\pm$3 \%		& 170.1$\pm$9.3 \\
\hline
\begin{minipage}{1.0in}
\begin{flushleft}
SVX single \\ tagged
\end{flushleft}
\end{minipage} 
		& 15 & 13$\pm$3 \%		& 178.1$\pm$7.9 \\
\hline
\begin{minipage}{1.0in}
\begin{flushleft}
SLT tagged \\ (no SVX tag)
\end{flushleft}
\end{minipage} 
		& 14 & 40$\pm$9 \%		& 142$\pm^{33}_{14}$ \\
\hline
\begin{minipage}{1.0in}
\begin{flushleft}
no tag (all jets \\ \ET $\ge$ 15 GeV)
\end{flushleft}
\end{minipage} 
		& 42 & 56$\pm$15 \%		& 181$\pm$9 \\
\hline
\end{tabular}
\end{center}
\end{table}
\noindent

\begin{figure}[t]
\begin{center}
\ifpdf
\includegraphics*[width=5in, height=4.5in, trim=0 0 -10 0]{d0_lepjet_mass}
\else
\psfig{figure=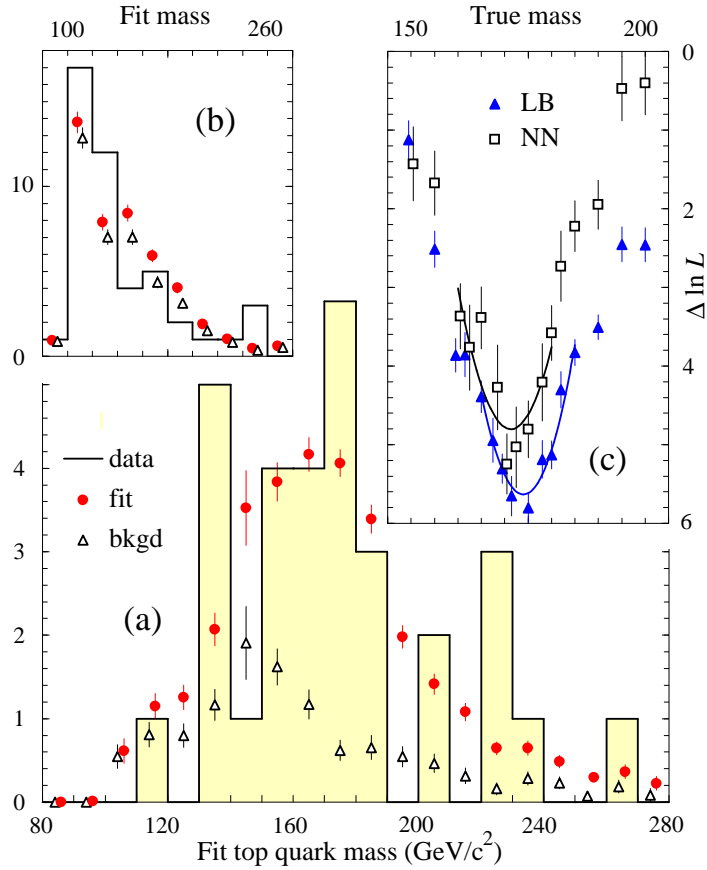,height=4.5in}
\fi
\caption{{\footnotesize D0 measurements of the top quark mass in the
lepton+jets events.}}
\end{center}
\end{figure}

\smallskip
D0 uses two multivariate discriminant analyses, LB-``low bias" and 
NN-``neural network", which use four variables to construct the top 
likelihood discriminant (D) to select
the top enriched and background enriched samples of events, which are the 
basis of D0 top mass and cross section analyses. 
The dominant systematic uncertainties (in GeV/c$^2$) are: jet energy 
measurement (4.0); background model (2.5); signal model (1.9);
fitting technique (1.5); calorimeter noise (1.3), yielding the total 
systematic error of 5.5 GeV/c$^2$.
A two-dimensional likelihood fit is performed in the M$_{fit}$ vs D plane.
A parabolic fit to the distribution of log(fit likelihood) vs M$_{fit}$
yields the result, M$_t$, corresponding to the minimum. Results of fits, plotted
in the signal-rich (a) and background-rich regions (b), are shown in 
Figure 2.
 
\noindent
The combined D0 result from the LB and NN
methods in the lepton+jets channel, 
with the correlations between the methods (88$\pm$4 \%) taken into account, is: 
\medskip
\begin{center}
	M$_t$=173.3$\pm$5.6(stat)$\pm$5.5(syst) GeV/c$^2$.
\end{center}
\medskip

\subsection{Mass Measurement in di-lepton channel}
\smallskip
In the di-lepton mode the situation is more complicated, as the problem is 
underconstrained (two missing neutrinos). Several techniques were developed.
All obtain a probability density distribution as a function of M$_t$, whose
shape allows identifying the most likely mass which satisfies a hypothesis
that a pair of top quarks were produced in an event, and that their decay 
products correspond to a given combination of leptons and jets. \MET~may, or
may not, be used. D0 developed two methods, the Neutrino Phase Space weighting
technique ($\nu$WT) and the Average Matrix Element technique (MWT), a modified
form of Dalitz-Goldstein~\cite{DG} and Kondo~\cite{DLM} methods. 
The combined result, from the $\nu$WT and MWT methods, is:
\medskip
\begin{center}
	M$_t$=168.4$\pm$12.3(stat)$\pm$3.6(syst) GeV/c$^2$.
\end{center}
\medskip
Three techniques of 
measurements of the top quark mass have been developed in CDF. Two use \MET 
(the ``neutrino
weighting" and the ``Minuit fitting" methods), one does not (a modification of
the Dalitz-Goldstein method,
which instead includes information about the parton distribution functions, 
transverse energy of the \TT system and angular correlations among the top decay
products in the definition of likelihood). The result obtained with the 
``neutrino weighting" method (essentially the D0 $\nu$WT) result is:
\medskip
\begin{center}
	M$_t$=167.4$\pm^{10.7}_{9.8}$(stat)$\pm$4.8(syst) GeV/c$^2$.
\end{center}
\medskip
\noindent
This result was available already last summer, and it was used in the 
CDF and CDF/D0
combined mass analyses. An analysis using the ``Minuit fitting" method yields:
\medskip
\begin{center}
	M$_t$=170.7$\pm$10.6(stat)$\pm$4.6(syst) GeV/c$^2$.
\end{center}
\medskip
\noindent
The Dalitz-Goldstein technique, which uses a single, ``best" combination of
leptons and jets in an event, gives:
\medskip
\begin{center}
	M$_t$=157.1$\pm$10.9(stat)$\pm^{4.4}_{3.7}$(syst) GeV/c$^2$.
\end{center}

\begin{table}[h]
\caption{\footnotesize Dominant systematic uncertainties in top mass measurements in the
dilepton mode in CDF(``neutrino weighting") and D0 (all errors in GeV/c$^2$).\label{tab:dilep}}
\begin{center}
\footnotesize
\begin{tabular}{|l|c|c|}
\hline
source of uncertainty		& CDF 	& D0	\\
\hline
jet energy scale		& 3.8	& 2.4 	\\
\hline
signal model (ISR,FSR)		& 2.8	& 1.8 	\\
\hline
Monte Carlo generators		& 0.6	& 0.0 	\\
\hline
background modelling		& 0.3	& 1.1 	\\
\hline
fitting technique		& 0.7	& 1.5 	\\
\hline
calorimeter noise		& 0.0	& 1.3	\\
\hline
total				& 4.8 	& 3.6 	\\
\hline
\end{tabular}
\end{center}
\end{table}

\subsection{Mass Measurement in all-jets channel}
\smallskip
Kinematical fits were performed in CDF  to a sample of events selected using
SVX tagging. The dominant errors are (in GeV/c$^2$):
jet energy scale (5.0); final state radiation (1.8); background model (1.7); Monte Carlo generators (0.8);
Monte Carlo statistics (0.6); initial state radiation (0.1); yielding the 
total systematic error of  5.7 GeV/c$^2$.
A parabolic fit to the likelihood distribution obtained from
fitting the data to a combination of signal and SM background templates yields:
\medskip
\begin{center}
	M$_t$=186.0$\pm$10.0(stat)$\pm$5.7(syst) GeV/c$^2$.
\end{center}

\section{Combined Top Mass Measurements}
\smallskip
The CDF (D0) mass measurements in three (two) channels are combined in each
of the experiments,
taking statistical uncertainties as uncorrelated. The systematic errors due
to the energy scale, signal model (ISR and FSR) and MC generator are taken
as 100\% correlated, and all other systematic errors are taken as
uncorrelated.

\begin{table}[h]
\caption{\footnotesize
Summary of the results used in the combined CDF, D0, and the joint CDF+D0
measurements of the top quark mass (all results in GeV/c$^2$).
\label{tab:combmass}}
\begin{center}
\footnotesize
\begin{tabular}{|l|c|c|}
\hline
channel			& CDF	 		& D0	\\
\hline
di-leptons		&167.4$\pm$10.3$\pm$4.8 & 168.4$\pm$12.3$\pm$3.6\\
\hline
lepton+jets		&176.1$\pm$4.8$\pm$5.3 & 173.3$\pm$5.6$\pm$5.5\\
\hline
all-jets		&186.0$\pm$10.0$\pm$5.7 & 	\\
\hline
combined		&176.1$\pm$4.0$\pm$5.1 & 172.1$\pm$5.2$\pm$4.9\\
\hline
\end{tabular}
\end{center}
\end{table}

\begin{figure}
\vspace{0.0cm}
\begin{center}
\ifpdf
\includegraphics*[width=3.5in, height=3.5in, trim=50 130 70 80]{mass_cdf_d0_bw}
\else
\psfig{figure=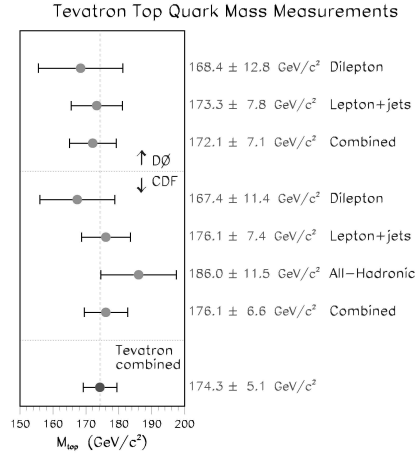,height=3.5in}
\fi
\caption{\footnotesize 
CDF and D0 measurements of the top quark mass using Tevatron Run-I
data.}
\end{center}
\end{figure}

\smallskip
The Tevatron (CDF+D0) average for Run-I was obtained
from the five CDF and D0 results in a similar manner to the way it was done 
to obtain the CDF and D0 averages. Systematic errors which 
do not depend directly on the Monte Carlo simulations 
(jet energy scale, backgrounds...) are 
taken as uncorrelated between the experiments, while those systematic 
errors which depend on the Monte Carlo model (ISR, FSR, PDF
dependence...) are treated as 100\% correlated between the experiments, since
both CDF and D0 rely on identical MC models. The result is:
\medskip
\begin{center}
	M$_t$=174.3$\pm$3.2(stat)$\pm$4.0(syst) GeV/c$^2$.
\end{center}
\section{\TT Pair Production Cross Section}

\begin{table}[h]
\caption{\footnotesize
CDF measurements of the \TT pair production cross section in individual 
channels,
together with the relevant values of acceptancies, trigger and tagging 
efficiencies, and the number of observed and expected backgrounds events.
\label{tab:cdfsigma}}
\begin{center}
\scriptsize
\begin{tabular}{|l|c|c|c|c|c|}
\hline
{}	&l+jets	& l+jets & di-leptons & all-jets & all-jets	\\
\hline
TAG type& SVX	& SLT	 & {}	      &	SVX	& double SVX 	\\
\hline
$\epsilon_{tagging}$ & 0.505$\pm$0.051 & 0.157$\pm$0.016 & 
& 0.544$\pm$0.057 & 0.17$\pm$0.05 \\
\hline
\begin{minipage}{1.0in}
\begin{flushleft}
geometrical and\\
kinematical cuts\\
acceptance
\end{flushleft}\end{minipage} 
&0.078$\pm$0.01&0.078$\pm$0.01&0.0074$\pm$0.0008&0.099$\pm$0.016&
0.263$\pm$0.045 \\
\hline
\begin{minipage}{1.0in}
\begin{flushleft}
trigger \\
acceptance
\end{flushleft}\end{minipage} 
& 0.90$\pm$0.07 & 0.90$\pm$0.07 & 0.98$\pm$0.01
& 0.998$\pm^{0.002}_{0.009}$ & 0.998$\pm^{0.002}_{0.009}$ \\
\hline
acceptance& 0.035$\pm$0.005 & 0.011$\pm$0.002 & 0.0074$\pm$0.0008
& 0.054$\pm$0.01 & 0.045$\pm$0.015 \\
\hline
number of events&	29	& 25	& 9	& 187	& 157	\\
\hline
background & 6.7$\pm$1.0 & 13.22$\pm$1.22 & 2.4$\pm$0.5 
& 144$\pm$12 & 120$\pm$18 \\
\hline
$\sigma_{tt}$ (in pb) & 5.1$\pm^{1.6}_{1.4}$ & 9.2$\pm^{4.8}_{3.9}$ 
& 8.2$\pm^{4.4}_{3.4}$ & 7.4$\pm^{3.8}_{3.1}$ & 7.8$\pm^{5.2}_{4.6}$ \\
\hline
\end{tabular}
\end{center}
\end{table}
\noindent
CDF combines the above cross section using a likelihood technique which takes
into account correlations in the uncertainties. Assuming the top quark mass of 
175 GeV/c$^2$ (in calculating all the corrections) the CDF value of the 
\TT pair production cross section is:
\begin{center}
	$\sigma_{tt}$=6.5$\pm^{1.7}_{1.4}$ pb 
\end{center}

\begin{figure}[h]
\vspace{0.5cm}
\begin{center}
\ifpdf
\includegraphics*[width=3.5in, height=3.5in, trim=70 150 50 125]{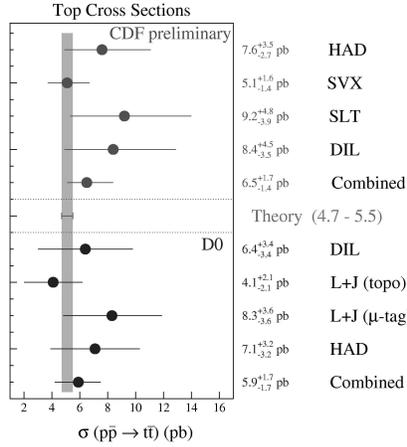}
\else
\psfig{figure=xs_cdf_d0_bw,height=3.5in}
\fi
\caption{{\footnotesize
CDF and D0 measurements of the top pair production cross section.  For comparison, the range of theoretical predictions~\cite{TH} 
for \TT pair production cross
section is also shown.}}
\end{center}
\end{figure}
 
\medskip\noindent
D0 measures the \TT cross section in 4 different samples. 

\begin{table}[h]
\begin{center}
\footnotesize
\begin{tabular}{|l|c|c|}
\hline
channel			&  events	& cross section (pb)\\
\hline
di-lepton + $e \nu$	& 9		& 6.4$\pm$3.3	\\
\hline
lepton+jets (topological)& 19	& 4.1$\pm$2.1\\
\hline
lepton+jets ($\mu$-tagged) & 11	& 8.3$\pm$3.5\\
\hline
all+jets 			& 41		& 7.1$\pm$3.2\\
\hline
\end{tabular}
\end{center}
\end{table}
\noindent
The D0 combined value (at M$_t$=172.1 GeV/c$^2$) is:
\medskip
\begin{center}
	$\sigma_{tt}$=5.9$\pm$1.7 pb 
\end{center}
\medskip
For comparison, the theoretical predictions~\cite{TH} 
for \TT pair production cross
section fall in the range of 4.7-5.5 pb, for M$_t$=175 GeV/c$^2$.

\section{Single Top Production}
\smallskip
Analysis of single top production offers a direct access to the Wtb vertex and should allow the measurement of the $|V_{tb}|$ element of Cabibbo-Kobayashi-Maskawa matrix. Anomalous couplings would lead to larger production rates, while the Standard Model cross section predictions are: 0.72$\pm$0.04 pb~\cite{Smith} and 1.70$\pm$0.20 pb~\cite{Stelzer}, for s-channel and t-channel processes, respectively.
Both D0 and CDF experiments conducted searches, although their sensitivity with the Run-I statistics is insufficient to detect signals of predicted magnitude. The CDF search for single top production was based on a fit to the  H$_T$ distribution for W+1,2,3 jet events, and assumed the Monte Carlo simulated shapes of H$_T$ distributions for QCD and \TT backgrounds. CDF finds a limit for the single top production cross section $\sigma$=13.5 pb at 95\% CL. D0 employed an array of neural nets to derive limits of  $\sigma$=17 pb at 95\% CL (s-channel) and  $\sigma$=22 pb at 95\% CL (t-channel).

\section{Standard Model consistency checks: Higgs boson mass  vs M$_t$}

\smallskip
The precision measurements of  electroweak parameters at LEP, SLC, FNAL and other precision experiments can be used to verify the consistency of the Standard Model and to infer bounds and constraints on its basic parameters. The leading-order top quark corrections are quadratic in top quark mass, M$_t$,  which allows quite precise ``determination" of M$_t$ $\it indirectly$ from other electroweak measurements. The dependence of the leading order corrections to the Higgs 
boson mass, M$_{H}$,  is logarithmic, and the bounds on Higgs mass are weaker with the current measurement errors. It is worthwhile to note that the value of M$_{top} \approx$ 175 GeV/c$^2$ could be obtained $\it indirectly$ from global fits to the electroweak parameters measured at  LEP, LEP-II, SLC and $\nu$ experiments but $\it only$ if one $\it assumes$ M$_{H} \approx$300 GeV/c$^2$. The fact that this particular value of M$_{H}$=300 GeV/c$^2$  was used in the electroweak fits - consistency checks of the Standard Model  from 1993-1996 was not emphasised when claims were made that LEP  ``measured" the top quark mass of about 175 GeV/c$^2$ in advance of  CDF and D0 direct measurements. It is also interesting to note that  a set of fits to the electroweak parameters in which both M$_t$ and M$_{H}$ are treated as free parameters were consistently pointing to a low Higgs mass (60-150 GeV/c$^2$) and a lower top quark mass (157-169 GeV/c$^2$). With all the excitement surrounding searches for a light Higgs at LEP-II the fact that the low mass Higgs would also point to the lower value of M$_t$ was not always remembered. With the precise measurements of M$_W$, M$_t$ and final results from LEP and LEP-II available, the most recent consistency checks of the Standard Model performed with global fits to all electroweak measurements  give poor fits. This could be an indication of the ``new physics"  beyond the Standard Model.  

\begin{table}[h]
\caption{\footnotesize
Results of the consistency checks of the Standard Model performed by LEP Electroweak Working Group~\cite{LepEWWG}. Indirect measurements of M$_t$ obtained using the global fits to electroweak parameters in years 1993-1996 assume M$_{H}$=300 GeV/c$^2$, and the second errors correspond to varying M$_{H}$ in the range 60-1000 GeV/c$^2$.
With more precise measurements available, starting in 1997 the global fits allow indirect determination of  both M$_t$ and M$_{H}$ (all masses in GeV/c$^2$). 
\label{tab:EWWG}}
\begin{center}
\footnotesize
\vspace{-0.2cm}
\begin{tabular}{|l|c|c|c|c|}
\hline
year		& {} &~~~~~~~~~ LEP ~~~~~~~&~~~~~~ all data~~~~~&M$_{H}$ (GeV/c$^2$)\\
\hline
1993 							&
\begin{minipage}{0.5in}
\begin{flushleft}
M$_t$ \\ $\chi^2$/NDF
\end{flushleft}\end{minipage} 			&
\begin{minipage}{0.5in}
\begin{center}
166$\pm^{17}_{19}\pm^{19}_{22}$ \\ 3.5/8
\end{center}\end{minipage} 			&
\begin{minipage}{0.5in}
\begin{center}
164$\pm^{16}_{17}\pm^{17}_{21}$ \\ 4.1/11
\end{center}\end{minipage} 			& 300\\
\hline
1994 							&
\begin{minipage}{0.5in}
\begin{flushleft}
M$_t$ \\ $\chi^2$/NDF
\end{flushleft}\end{minipage} 			&
\begin{minipage}{0.5in}
\begin{center}
173$\pm^{12}_{13}\pm^{18}_{20}$ \\ 7.6/9
\end{center}\end{minipage} 			&
\begin{minipage}{0.5in}
\begin{center}
171$\pm^{11}_{12}\pm^{18}_{19}$ \\ 4.4/11
\end{center}\end{minipage} 			& 300\\
\hline
1995 							&
\begin{minipage}{0.5in}
\begin{flushleft}
M$_t$ \\ $\chi^2$/NDF
\end{flushleft}\end{minipage} 			&
\begin{minipage}{0.5in}
\begin{center}
170$\pm{10}\pm^{17}_{29}$ \\ 18/9
\end{center}\end{minipage} 			&
\begin{minipage}{0.5in}
\begin{center}
178$\pm{8}\pm^{17}_{20}$ \\ 28/14
\end{center}\end{minipage} 			& 300 \\
\hline
1996 							&
\begin{minipage}{0.5in}
\begin{flushleft}
M$_t$ \\ $\chi^2$/NDF
\end{flushleft}\end{minipage} 			&
\begin{minipage}{0.5in}
\begin{center}
171$\pm{8}\pm^{17}_{19}$ \\ 10/9
\end{center}\end{minipage} 			&
\begin{minipage}{0.5in}
\begin{center}
177$\pm{7}\pm^{16}_{19}$ \\ 24/14
\end{center}\end{minipage} 			& 300\\
\hline
year		& {} &~~~~~~~~~~~ LEP ~~~~~~~~ & ~~~~~~~~~all data~~~~~~~~&~~ exclude M$_t$, M$_W$ ~\\
\hline
1997 							&
\begin{minipage}{0.5in}
\begin{flushleft}
M$_t$ \\ M$_{H}$\\ $\chi^2$/NDF
\end{flushleft}\end{minipage} 			&
\begin{minipage}{0.5in}
\begin{center}
158$\pm^{14}_{11}$ \\  83$\pm^{168}_{49}$ \\ 8/9
\end{center}\end{minipage} 			&
\begin{minipage}{0.5in}
\begin{center}
173.1$\pm{5.4}$ \\ 115$\pm^{116}_{66}$ \\ 17/15
\end{center}\end{minipage} 			& 
\begin{minipage}{0.5in}
\begin{center}
157$\pm^{10}_{9}$ \\ 41$\pm^{64}_{21}$ \\ 14/12
\end{center}\end{minipage} 			\\
\hline
1998							&
\begin{minipage}{0.5in}
\begin{flushleft}
M$_t$ \\ M$_{H}$\\ $\chi^2$/NDF
\end{flushleft}\end{minipage} 			&
\begin{minipage}{0.5in}
\begin{center}
160$\pm^{13}_{9}$ \\  60$\pm^{127}_{35}$ \\ 4/9
\end{center}\end{minipage} 			&
\begin{minipage}{0.5in}
\begin{center}
171.1$\pm{4.9}$ \\ 76$\pm^{85}_{47}$ \\ 15/15
\end{center}\end{minipage} 			& 
\begin{minipage}{0.5in}
\begin{center}
158$\pm^{9}_{8}$ \\ 32$\pm^{41}_{15}$ \\ 13/12
\end{center}\end{minipage} 			\\
\hline
1999 							&
\begin{minipage}{0.5in}
\begin{flushleft}
M$_t$ \\ M$_{H}$\\ $\chi^2$/NDF
\end{flushleft}\end{minipage} 			&
\begin{minipage}{0.5in}
\begin{center}
172$\pm^{11}_{11}$ \\  134$\pm^{268}_{81}$ \\ 11/9
\end{center}\end{minipage} 			&
\begin{minipage}{0.5in}
\begin{center}
173.2$\pm{4.5}$ \\ 77$\pm^{69}_{39}$ \\ 23/15
\end{center}\end{minipage} 			& 
\begin{minipage}{0.5in}
\begin{center}
167$\pm^{11}_{8}$ \\ 55$\pm^{84}_{27}$ \\ 21/12
\end{center}\end{minipage} 			\\
\hline
2000 							&
\begin{minipage}{0.5in}
\begin{flushleft}
M$_t$ \\ M$_{H}$\\ $\chi^2$/NDF
\end{flushleft}\end{minipage} 			&
\begin{minipage}{0.5in}
\begin{center}
179$\pm^{13}_{10}$ \\  135$\pm^{262}_{83}$ \\ 13/9
\end{center}\end{minipage} 			&
\begin{minipage}{0.5in}
\begin{center}
174.3$\pm^{4.4}_{4.1}$ \\ 60$\pm^{52}_{29}$ \\ 21/15
\end{center}\end{minipage} 			& 
\begin{minipage}{0.5in}
\begin{center}
169$\pm^{10}_{8}$ \\ 56$\pm^{75}_{27}$ \\ 19/12
\end{center}\end{minipage} 			\\
\hline
2001 							&
\begin{minipage}{0.5in}
\begin{flushleft}
M$_t$ \\ M$_{H}$\\ $\chi^2$/NDF
\end{flushleft}\end{minipage} 			&
\begin{minipage}{0.5in}
\begin{center}
186$\pm^{13}_{11}$ \\  260$\pm^{404}_{155}$ \\ 15.5/8
\end{center}\end{minipage} 			&
\begin{minipage}{0.5in}
\begin{center}
175.8$\pm^{4.4}_{4.3}$ \\ 88$\pm^{53}_{35}$ \\ 22.9/15
\end{center}\end{minipage} 			& 
\begin{minipage}{0.5in}
\begin{center}
169$\pm^{12}_{9}$ \\ 81$\pm^{109}_{40}$ \\ 18.9/12
\end{center}\end{minipage} 			\\
\hline
year		& {} &~~~~~~~~~~~ LEP ~~~~~~~~ & ~~~~~~~~~all data~~~~~~~~&~~ exclude M$_t$ ~\\
\hline
2002							&
\begin{minipage}{0.5in}
\begin{flushleft}
M$_t$ \\ M$_{H}$\\ $\chi^2$/NDF
\end{flushleft}\end{minipage} 			&
\begin{minipage}{0.5in}
\begin{center}
184$\pm^{13}_{11}$ \\  228$\pm^{367}_{136}$ \\ 13.3/9
\end{center}\end{minipage} 			&
\begin{minipage}{0.5in}
\begin{center}
174.3$\pm^{4.5}_{4.3}$ \\ 81$\pm^{52}_{33}$ \\ 29.7/15
\end{center}\end{minipage} 			& 
\begin{minipage}{0.5in}
\begin{center}
180$\pm^{11}_{9}$ \\ 117$\pm^{161}_{63}$ \\ 17.9/12
\end{center}\end{minipage} 			\\
\hline
\end{tabular}
\end{center}
\end{table}

\section{Prospects for Run-II. Is it only top ?}
\smallskip
In Run-IIa, which started at the end of 2001, CDF and D0 are each expected to collect 2 fb$^{-1}$ of integrated luminosity. With the new Main Injector, the $\rm p\bar{\rm p}$ collisions take place at $\sqrt{\rm s}$ = 1.96 TeV, and the \TT cross section is $\approx$35\% larger than at Run-I. Because of different beam crossing time (396 ns and 132 ns later, instead of 3.5 $\mu$s in Run-I) the number of multiple interactions per event will be less than in Run-I. CDF has a new calorimeter with a much better energy resolution in the pseudorapidity range 1.1$< |\eta| <$3.5, and a new SVX with double the Run-I tagging efficiency. CDF also added a time-of-flight system and its muon coverage has been doubled to cover the range $|\eta|<$2. D0 has a new SVX to allow better b-tagging, and has added a solenoid to allow momentum reconstruction for charged particles. D0 has excellent lepton ($|\eta|<$2 for muons, $|\eta|<$2.5 for electrons) and tracking coverage ($|\eta|<$3). 
With the increased integrated luminosity (20x), combined with improvements to CDF and D0 detectors
and larger \TT cross section, the number of reconstructed top events will increase by a factor of
$\approx$20-70, depending on the final state and tagging requirements. Both experiments estimate that they'll measure the top quark mass with an error of $\Delta M_{top}$= 2-3 GeV/c$^2$ (compared with 7 GeV/c$^2$ in Run-I) and the \TT cross section with an uncertainty of about 8\% (about 15\% in Run-I). The biggest challenge for both experiments will be to reducing the systematic errors to take full advantage of expected large statistics. Analysis of single top production offers a direct access to the Wtb vertex and should allow the measurement of the $|V_{tb}|$ element of Cabibbo-Kobayashi-Maskawa matrix. Anomalous couplings would lead to anomalous angular distributions and larger production rates, while the expected cross sections are of the order of 1-2 pb.

\begin{table}[h]
\caption{\footnotesize
Prospects for CDF and D0 experiments in Tevatron Run-II, compared with the corresponding values of selected characteristics  from Run-I. \label{tab:prospects}}
\begin{center}
\footnotesize
\begin{tabular}{|l|c|c|c|}
\hline
{}		& RUN-I & RUN-IIa CDF & RUN-IIa D0\\
\hline
``typical"  luminosity  (cm$^{-2}s^{-1}$)
		& 1.6$\times 10^{30}$ & 8.6$\times 10^{31}$ & 8.6$\times 10^{31}$\\ 
\hline
integrated  luminosity
		& ~110 pb$^{-1}$ & 2 fb$^{-1}$	& 2 fb$^{-1}$ \\
\hline
dilepton events
		& ~10/exp 	& 	140			& 200 \\
\hline
lepton+$\ge$4 jets  
		& ~20/exp 	& 1500			& 1800 \\
\hline
lepton+$\ge$3 jets+$\ge$ 1b jet tag
		& ~30/exp 	& 1400			& 1400 \\
\hline
lepton+$\ge$4 jets+$\ge$ 2b jet tag
		& ~5/exp 	& 610			& 450 \\
\hline
$\Delta$M$_t$ 
		& 7 GeV/c$^2$ 	& 2-3 GeV/c$^2$		& 2-3 GeV/c$^2$ \\
\hline
$\Delta\sigma$(\TT)
		& ~30\% 			&  ~8\%				& ~8\% \\
\hline
\end{tabular}
\end{center}
\end{table}
\vspace{-0.3cm}
Perhaps even more importantly, the \TT  and single top events constitute background to any new physics. As a consequence of the large top mass, the event selection cuts in
top analyses are virtually identical to those applied in many analyses looking 
for physics beyond the SM 
(Supersymmetry, Technicolor, et cetera...). The measured \TT cross section values
depend on the top quark mass, whose value has been determined in CDF and D0 using 
various kinematical fitting techniques $\it and$ the assumption that 
events are just 
the \TT events and the SM background. If the sample is not not exclusively due to the \TT events and the SM background, the mass
measurements may be incorrect. 
If an additional process were present,  
the number of observed events would not agree then
with the MC predictions obtained with the measured value of M$_t$. 
It is thus imperative to
compare various distributions of the reconstructed top quarks, and 
especially those
of the \TT-system, with the SM predictions. Discrepancies could indicate new
physics. 
Both CDF and D0 made numerous comparisons. No significant disagreements
were found, as perhaps should be expected given the still limited statistics.
However, there exist a few hints that the simplest hypothesis (that the top 
candidate events are just the \TT events and SM background) may not be
entirely correct. With the luminosity of 2 fb$^{-1}$ per experiment they should be monitored carefully, as  they may be offering us glimpses of new physics.

\vfill\noindent
{ i.} CDF \TT cross section seems a little high compared to the theoretical 
predictions; it would be more consistent with the lower value of M$_t$; the
indirect measurements of M$_t$, based on the global checks of  the SM 
{\it {excluding}} the direct M$_t$ measurements, alo prefer lower M$_t$ 
($\approx$ 157-169 GeV/c$^2$)
\medskip
\vfill\noindent
{ ii.} There is an excess of W+2jet and W+3jet events (13 where 4.4$\pm 0.6$ are expected) with double tagged jets (tagged both with SVX and SLT) in the tagged jet
multiplicity distribution in the CDF. In addition, the kinematical properties of those events don't agree well with the SM predictions~\cite{multitagged}.

\begin{figure}[h]
\begin{center}
\ifpdf
\includegraphics*[width=4.5in, height=4.5in, trim=0 10 10 10]{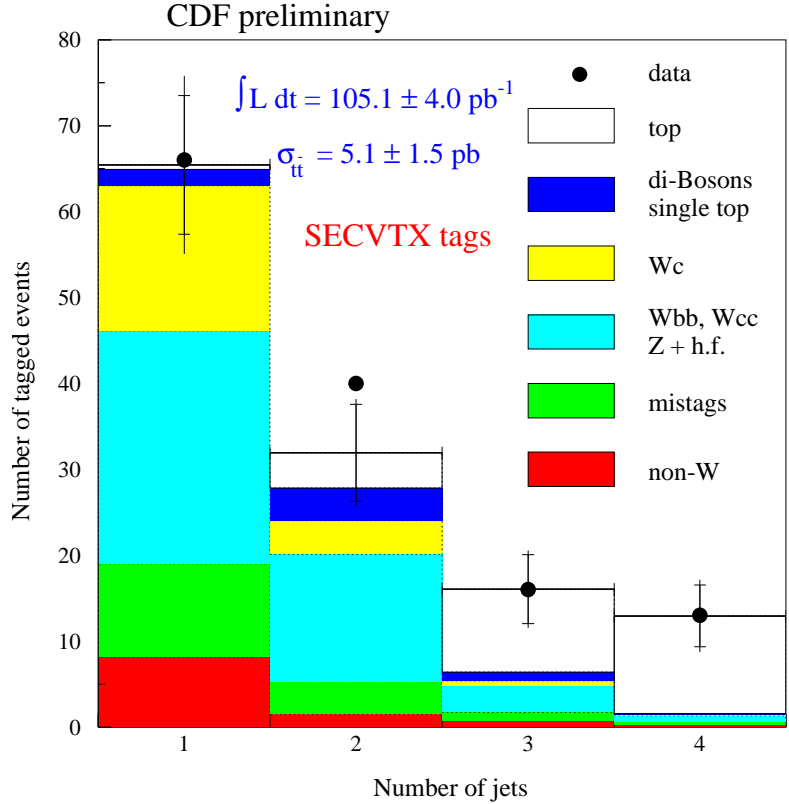}
\else
\psfig{figure=new_secvtx_tt,height=4.5in}
\fi
\caption{{\footnotesize
Number of W+Njets events as a function of the number of jets, N, for CDF top candidates with at least one of jets tagged with SVX, the CDF vertex detector. The excess in W+2jet events is   more pronounced if  jets are tagged both with SVX and SLT.}}
\end{center}
\end{figure}

\vfill\noindent
{ iii.} There is a hint of an increase of the
reconstructed top quark mass with the number of jets in an event.
\medskip
\vfill\noindent
{ iv.} Two (out of 9) CDF di-lepton events poorly fit  the \TT hypothesis and have unexpectedly large
\MET+$\Sigma E_t^{lepton}$. One such 
event exists in the D0 sample.
\medskip
\vfill\noindent
{ v.} The distributions of the \TT mass, in both CDF and D0, seem to have a few more
events than expected in the high mass region.
\medskip
\vfill\noindent
{ vi.} The transverse momentum distribution of the \TT system for the sample of 
32 CDF tagged lepton+jets events, 
seems a little harder than
expected, based on the Monte Carlo calculations. D0 data does not show any deviations from SM expectations.
\medskip
\vfill\noindent
{ vii.} The rapidity distribution
(Figure 6)
of the \TT system for the sample of 32 CDF tagged lepton+jets events has a
strikingly different shape than that based on MC simulations.  The rapidity  variable probes directly the {\it fitted} longitudinal component of the neutrino momenta and, as such, is perhaps more sensitive than other variables to the correctness of the original hypothesis that the fitted events are the \TT events.

\begin{figure}[h]
\vspace{0.7cm}
\begin{center}
\ifpdf
\includegraphics*[width=4.5in, height=4.5in, trim=0 0 0 0]{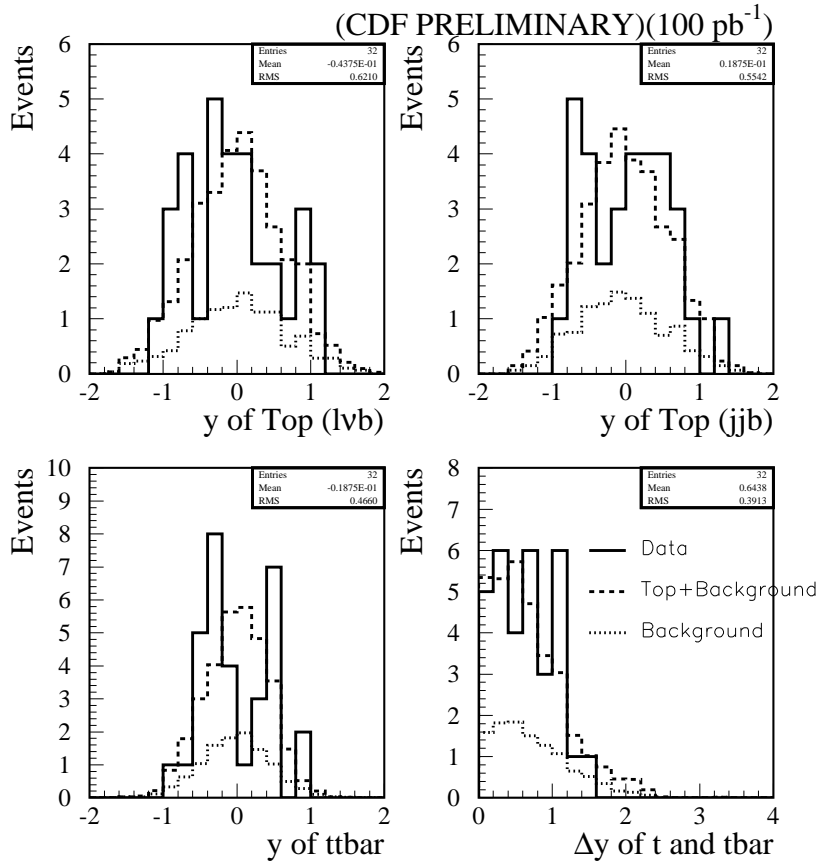}
\else
\psfig{figure=cdf_tt_y,height=4.5in}
\fi
\caption{{\footnotesize
CDF distributions of the rapidity for top, antitop and the \TT system of the fitted top quarks in the sample of 32 tagged lepton+jets events.}}
\end{center}
\end{figure}
\medskip
\vfill\noindent
However, the D0
pseudorapidity plot
(Figure 7)
is in good agreement with expectations~\cite{Mumbai}.

\begin{figure}[h]
\begin{center}
\ifpdf
\includegraphics*[width=4.5in, height=4.5in, trim=0 10 10 10]{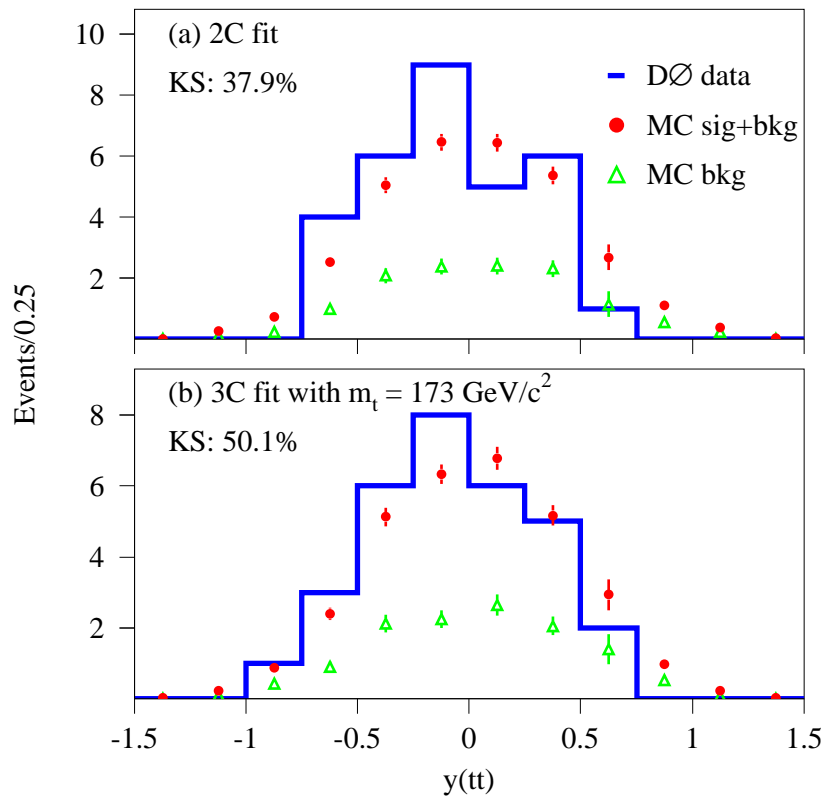}
\else
\psfig{figure=y_tt,height=4.5in}
\fi 
\caption{{\footnotesize
D0 distribution of the rapidity of \TT system in the
lepton+jets events.}}
\end{center}
\end{figure}

\section*{Acknowledgments}
\smallskip
I would like to thank the Conference
Organizers for their very warm hospitality and for making the Epiphany 2003
an exceptionally well organized conference. It was an especially  memorable experience for me;  I still vividly remember my times as a student, and later an assistant, at the Institute of Physics of Jagiellonian University where the conference was held.


\begin{thebibliography}{99}
\bibitem{CDF94}F. Abe {\it et al}, \Journal{\PRL}{73}{225}{1994}.

\bibitem{CDF95}F. Abe {\it et al}, \Journal{\PRL}{74}{2626}{1995}.

\bibitem{D095}S. Abachi {\it et al}, \Journal{\PRL}{74}{2632}{1995}.

\bibitem{Giele}F.A. Berends, H. Kuijf, B. Tausk, W.T. Giele,
\Journal{\NPB}{37}{32}{1991} (both CDF and D0 experiments used the  VECBOS W+3jets calculation, with an additional jet added in the fragmentation process simulated using HERWIG, rather than the more appropriate VECBOS W+4jets calculation because of prohibitive amount of CPU required by the latter.)

\bibitem{D0sear}S. Abachi {\it et al}, \Journal{\PRL}{79}{1203}{1997},
S. Abachi {\it et al}, \Journal{\PRD}{58}{052001}{1998}.

\bibitem{CDFsear}F. Abe {\it et al}, \Journal{\PRL}{79}{3585}{1997}.
F. Abe {\it et al}, \Journal{\PRL}{80}{2773}{1998}.

\bibitem{DG}Gary R. Goldstein and R.H. Dalitz, \Journal{\PRD}{45}
{1531}{1992};Gary R. Goldstein, K. Sliwa and R.H. Dalitz, \Journal{\PRD}
{47}{967}{1993}.

\bibitem{DLM}K. Kondo {\it et al.} J. Phys. Soc. Japan. {\bf 62}
(1993) 1177.

\bibitem{TH} E. Laenen, J. Smith and W.L. van Neerven, Journal{\PLB}{321}{251}
{1994}; E. Berger and H. Contapanagos, \Journal{\PLB}{361}{115}{1995},
\Journal{\PRD}{54}{3085}{1996}; S. Catani, M.L. Mangano, P. Nason and L. Trentadue,
\Journal{\PLB}{378}{329}{1996}.
 

\bibitem{Smith} M.C. Smith, S. Willenbrock, \Journal{\PRD}{54}
{6696}{1996}

\bibitem{Stelzer} T. Stelzer, Z. Sullivan, S. Willenbrock, \Journal{\PRD}{56}
{5919}{1997}.


\bibitem {LepEWWG} LEP Electroweak Working Group Reports, 1993-2002, CERN.


\bibitem{multitagged}D.Acosta {\it et al}, \Journal{\PRD}{65}{072005}{2002}.

\bibitem{Mumbai} ``Top mass and cross section results from CDF and D0 at the Fermilab
Tevatron", K. Sliwa, invited talk at 13th Topical Conference on Hadron 
Collider Physics, 
Tata Institute of Fundamental Research, Mumbai, India, January 14-20, 1999;
in Proceedings, p 169; World Scientific, 1999.




\end{thebibliography}
\end{document}